\documentclass[10pt]{article}
\usepackage{amsmath}
\usepackage{graphicx}
\usepackage{caption}
\usepackage{subcaption}
\usepackage{color}
\usepackage{orcidlink}
\begin{document}
\baselineskip=16pt

\begin{center}
{\large {\bf Effects of external field and potential on non-relativistic quantum particles in disclinations background }}   
\end{center}

\vspace{0.3cm}
\begin{center}
{\bf Faizuddin Ahmed\orcidlink{0000-0003-2196-9622}}\footnote{\bf faizuddinahmed15@gmail.com}\\
\vspace{0.1cm}
{\it Department of Physics, University of Science \& Technology Meghalaya, Ri-Bhoi, 793101, India.}\\
\vspace{0.3cm}
{\bf Allan R. P. Moreira\orcidlink{0000-0002-6535-493X}}\footnote{\bf allan.moreira@fisica.ufc.br}\\
\vspace{0.1cm}
{\it Secretaria da Educa\c{c}\~{a}o do Cear\'{a} (SEDUC), Coordenadoria Regional de Desenvolvimento da Educa\c{c}\~{a}o (CREDE 9),  Horizonte, Cear\'{a}, 62880-384, Brazil.}\\
\end{center}

\vspace{0.3cm}

\begin{abstract}

In this work, we investigate the behavior of non-relativistic quantum particles immersed in a cosmic string space-time background. Our study involves the examination of these particles as they interact with a range of influences, including potential, magnetic, and quantum flux fields. We employ analytical methods to solve the associated wave equation, leading to the derivation of eigenvalue solutions for this quantum system. Subsequently, we leverage these eigenvalue solutions to scrutinize several potential models. For each model, we present and engage in a thorough discussion of the corresponding eigenvalue solutions. In an extension of our investigation, we explore the thermodynamic and magnetic properties of the quantum system when it is exposed to non-zero temperature conditions, denoted by $T \neq 0$. Our analysis encompasses the calculation of essential parameters such as the partition function for the system and other pertinent functions. Following these calculations, we meticulously examine and interpret the outcomes, shedding light on the system's behavior and characteristics in the presence of temperature variations. Furthermore, we calculate entropic information to investigate the location of particles in the system.

\end{abstract}

\vspace{0.1cm}

{\bf keywords:} Non-Relativistic wave equation: Schr\"{o}dinger equation; Linear defects: disclinations; Solutions of wave equations; Special functions; Thermodynamic properties;  Magnetic properties; Quantum information.

\vspace{0.1cm}


\section{Introduction}
\label{sec1}

The eigenvalue solutions stemming from the wave equation within potential models play a crucial role in describing a wide array of physical systems, carrying significant importance in multiple domains of physics and chemistry. However, only a limited set of potential models are recognized for their ability to provide exact solutions to the wave equation. Among these well-known models are the Coulomb potential, which is fundamental in understanding the hydrogen atom, and the harmonic oscillator problem, extensively studied and documented in \cite{bb17, bb18, bb20}. Additionally, the pseudoharmonic potential, as explored in \cite{SMI, SMI3, fa, fa2, fa3, VK, fa4, fa10}, and the Mie potentials as discussed in \cite{fa4, fa5, fa6, fa9} have been identified as potential models that yield exact solutions. Furthermore, researchers have extended their investigations to encompass various other potential models, such as the Yukawa potential, Hulthén potential, Eckart potential, Morse potential, Manning-Rosen potential, Deng-Fan potential, and Poschl-Teller potential, among others. While these models may not offer exact solutions, they have been employed to study the wave equation and generate approximate eigenvalue solutions, both in two-dimensional (2D) and three-dimensional (3D) contexts, often employing suitable approximation schemes.

The scientific literature is replete with investigations into quantum mechanical phenomena involving topological defects known as cosmic strings. These studies encompass a wide range of scenarios, both in the presence and absence of magnetic and quantum flux fields, as well as the influence of various potentials. These investigations are categorized into two main regimes: Relativistic and non-relativistic regimes. In the relativistic limit, research endeavors are carried out and can be found in \cite{ERFM, ABHA, ABHA2, ALCO, EPL, CJP}. In the non-relativistic regime, research has been primarily focused on several key aspects: quantum dynamics of particles within the cosmic string space-time background \cite{CF}, quantum particles interacting with a diverse array of potentials, including \cite{VBB, GAM}, particles interacting with specific potentials such as the Coulomb and harmonic oscillator potentials \cite{JLAC}, quantum systems involving neutrons within a mean-field Woods-Saxon potential \cite{KJ}, studies exploring the confinement effects of Aharonov-Bohm (AB) flux and magnetic fields, typically modeled using a screened modified Kratzer potential \cite{COE}, research involving particles interacting with a Coulomb ring-shaped potential \cite{ZW}, investigations centered around particles confined by AB-flux and magnetic fields, often employing a Yukawa potential \cite{COE2}, quantum motions of particles influenced by a trigonometric non-central potential \cite{CC}, studies involving particles subjected to a pseudo-Coulomb potential in combination with an improved ring-shaped potential \cite{ANI2}, quantum motions of particles within Pöschl-Teller double-ring-shaped Coulomb and double-ring-shaped oscillator potentials \cite{AA2}, quantum motions of particles within a doubled ring-shaped Coulomb oscillator potential \cite{CC2}, investigations exploring solutions of the d-dimensional time-independent cosmic string equation for non-central potentials, including the Rosen-Morse, Scarf II, and Scarf I potentials \cite{CC3}, research dealing with Hyperbolic Scarf potentials combined with Pöschl-Teller and Manning-Rosen potentials \cite{CC4}, quantum particles undergoing interactions with a magnetic field in the space-time of a spinning cosmic string \cite{AA3}. Collectively, these comprehensive investigations contribute significantly to our understanding of the intricate quantum behavior of particles in the presence of cosmic strings and their interactions with various external influences, thus advancing our knowledge of a wide range of physical phenomena.

Information theories are gaining a lot of attention these days \cite{Wang:2023xhy,Santana-Carrillo:2022gmu,Gil-Barrera:2022ayj,Lima:2022dts,Lima1,Moreira:2023fui,Inyang:2023din,Ren:2023oxg,Benabdallah:2023gku,Lopez-Garcia:2023dah, Mondal:2023wmf}. The main theory most used is information entropy proposed by Shannon in 1948 \cite{Shannon}. This is because this theory can be applied to quantum systems to investigate the location conditions of particles through the system's probability densities. Shannon derives this theory seeking to find the best way to transmit a message between a source and its receiver (least noise or interference). Furthermore, an uncertainty relation is defined through entropic information, which gives us an alternative to Heisenberg uncertainty principle \cite{Beckner, Bialy}.

Our primary goal is to explore the behavior of non-relativistic particles within the framework of cosmic disclinations. We take into account the coexistence of magnetic and quantum flux fields along with an external potential. Our specific focus lies in showcasing how the presence of cosmic dislocations, as well as the magnetic and quantum flux fields, significantly shape the eigenvalue solutions of the system. Furthermore, we extend our investigation to analyze the thermodynamic and magnetic properties of this system when it operates under the influence of finite temperature conditions ($T \neq 0$). In doing so, we aim to thoroughly examine the consequences of these factors on the overall characteristics of the system, providing a comprehensive understanding of its behavior. Furthermore, we calculate the entropy information to investigate the location conditions of the wave eigenfunctions.

This paper follows a structured outline, which is as follows: Section \ref{sec2}: We begin by delving into the Schrödinger equation within the context of disclinations geometry. Here, we take into account the presence of both uniform magnetic and quantum flux fields, in addition to an external potential. Our focus lies in solving the resulting radial equation, employing the Nikiforov-Uvarov (NU) method, with a particular emphasis on the anharmonic oscillator potential. Moving forward, we take this eigenvalue solution and apply it to a variety of molecular potential models in subsection \ref{sec2.1}. In section \ref{sec3}, we perform calculations to determine the partition function, employing the derived energy expression. This analysis leads us to explore the thermodynamic properties of the system. Furthermore, we shift our attention to the magnetic properties of the quantum system, investigating them in detail. In section \ref{sec4}, we calculate the entropic information (Shannon entropy) and generate a few tables by settings different parameters. In section \ref{sec5}, we draw our paper to a conclusion in this section, summarizing our findings and providing closing remarks. Throughout the analysis, the system of units is chosen where $c=1=\hbar=G$.   

\section{Quantum System with external fields and potential in conical geometry}
\label{sec2}

In this section, we aim to investigate non-relativistic quantum particles that find themselves confined by an Aharonov-Bohm flux field, in conjunction with an external potential within the backdrop of cosmic disclinations. We commence this exploration by considering a static and cylindrical symmetric metric, which characterizes the cosmic string space-time. This metric in four-dimensions can be represented as: $ds^2=-dt^2+dr^2+\alpha^2\,r^2\,d\phi^2+dz^2$. Here, the parameter $\alpha$ is constrained within the range $0 < \alpha < 1$. This parameter serves as a crucial descriptor of the topological defect, signifying positive curvature within the cosmic space-time. Notably, the coordinate $\phi$ spans from $0 \leq \phi < 2\pi$, albeit with a modified periodicity of $2\pi\alpha$ around the $z$-axis. The spatial aspect of this metric, $ds^2_{3D}=dr^2+\alpha^2\,r^2\,d\phi^2+dz^2=g_{ij}\,dx^i\,dx^j$, embodies the cosmic disclinations inherent in this geometry. Here, the indices $i$ and $j$ take values from 1 to 3, with specific components of the metric tensor, namely, $g_{11}$ ($=1$), $g_{22}$ ($=\alpha^2\,r^2$), and $g_{33}$ ($=1$), dictating the spatial relationships within this cosmic disclinations. This distinctive space-time setting provides the foundation for our exploration of quantum dynamics within the influence of cosmic dislocations and associated fields. 

The time-independent Schr\"{o}dinger wave equation with potential $V(r)$ is described by the following wave equation \cite{bb17, bb18, bb20,SMI, SMI3, fa, fa2, fa3, VK, fa4, fa5, fa6, fa9, fa10}
\begin{eqnarray}
\Big[-\frac{1}{2\,M}\frac{1}{\sqrt{g}}D_{i}\Big(\sqrt{g}g^{ij}D_{j}\Big)+V(r)\Big]\Psi=E \Psi,
\label{2}
\end{eqnarray}
where $E$ is the energy of the particles, $g$ is the determinant of the metric tensor $g_{ij}$ with $g^{ij}$ its inverse, $D_i=(\partial_i-i\,e\,A_i)$ with $e$ is the electric charges and $A_i$ is the electromagnetic three-vector potential. Here $\Psi (r, \phi, z)$ is the wave function and in terms of a radial wave function $\psi (r)$, it can be expressed as: $\Psi(r, \phi, z)=e^{i\,\ell\,\phi}\,e^{i\,k\,z}\,\psi (r)$, where $\ell$ is the orbital quantum number and $k$ is an arbitrary constant (choosing $k=0$ here for simplicity). Furthermore, the electromagnetic three-vector potential $\vec{A}=(A_1+A_2)\,\hat{\phi}$, where $\vec{A_1}=-\frac{1}{2\,\alpha}\,(\vec{B}\times \vec{r})$ and $\vec{A}_2=\Big(\frac{\Phi_{AB}}{2\,\pi\,\alpha\,r}\Big)\,\hat{\phi}$. Here $\Phi_{AB}=\Phi\,\Phi_0=const$ is called the Aharonov-Bohm quantum flux, and $\Phi_0=2\,\pi\,e^{-1}$. The uniform magnetic field will be $\vec{B}=\vec{\nabla} \times \vec{A}_1=-B_0\,\hat{z}$ which is perpendicular to the plane where the particles are confined, and $\vec{\nabla} \cdot \vec{A}_2=0$.

Therefore, expressing the wave equation (\ref{2}) in the considered geometry background $ds^2_{3D}=g_{ij}\,dx^i\,dx^j$, and using the electromagnetic three-vector potential and the wave function ansatz, we obtain the following differential equation 
\begin{eqnarray}
\psi''(r)+\frac{1}{r}\,\psi'(r)+\Big[2\,M\,(E-V)-\frac{1}{\alpha^2\,r^2}\,(\ell+M\,\omega_c\,r^2-\Phi)^2\Big]\,\psi(r)=0.
\label{3}
\end{eqnarray}
That can be written as 
\begin{eqnarray}
\psi''(r)+\frac{1}{r}\,\psi'(r)+\Bigg[2\,M\,(E-V(r))-\frac{\ell'^2}{r^2}-\frac{M^2\,\omega^2_{c}}{\alpha^2}\,r^2-\frac{2\,M\,\omega_c}{\alpha}\,\ell'\Bigg]\,\psi(r)=0,
\label{4}
\end{eqnarray}
where $\ell'=\frac{1}{\alpha}\,|\ell-\Phi|$ is the effective orbital quantum number, and $\omega_c=\frac{|e|\,B}{2\,M}$ is the cyclotron frequency.

The effective potential of the system (by changing the function $\psi(r)=\frac{R(r)}{r}$ into the Eq. (\ref{4}), one can find the one-dimensional Schr\"{o}dinger-like wave equation) is given by
\begin{eqnarray}
V_{eff} (r)=V(r)+\frac{(\ell-\Phi)^2}{2\,M\,\alpha^2\,r^2}+\frac{e^2\,B^2}{8\,M\,\alpha^2}\,r^2+\frac{|e|\,B}{2\,M}\,\frac{|\ell-\Phi|}{\alpha^2}.
\label{5}
\end{eqnarray}

In this analysis, we are interested on the following potential $V (r)$ called anharmonic oscillator or pseudoharmonic-type potential given by
\begin{eqnarray}
V(r)=a\,r^2+\frac{b}{r^2}+c,
\label{6}
\end{eqnarray}
where $a, b, c$ characterise the potential parameters. For $b \to 0$, $c \to 0$, we have a harmonic oscillator potential. Whereas for $a \to 0$, $c \to 0$, one will have an inverse square potential. This type of potential has been used to study the relativistic and non-relativistic quantum systems by several authors in the literature (see, for examples \cite{SMI,SMI3,fa,fa2,fa3,VK,fa4}). 

With this potential (\ref{6}), the effective potential of the quantum system becomes
\begin{eqnarray}
V_{eff}(r)=\Big(a+\frac{e^2\,B^2}{8\,M\,\alpha^2}\Big)\,r^2+\Big(b+\frac{(l-\Phi)^2}{2\,M\,\alpha^2}\Big)\,\frac{1}{r^2}+\frac{|e|\,B}{2\,M}\,\frac{(l-\Phi)}{\alpha^2}+c.
\label{pp}
\end{eqnarray}
One can see that the effective potential of the quantum system depends on the topological defects characterised by the parameter $\alpha$, and the magnetic field $B$, and the quantum flux $\Phi_{AB}$.

Thereby, substituting this potential $V(r)$ into the Eq. (\ref{4}), we obtain the following radial equation:
\begin{equation}
\psi''(r)+\frac{1}{r}\,\psi'(r)+\Big(\Lambda-\frac{j^2}{r^2}-\Omega^2\,r^2\Big)\,\psi(r)=0,
\label{8}
\end{equation}
where we have set the parameters
\begin{eqnarray}
\Lambda=2\,M\,(E-c)-\frac{2\,M\,\omega_c}{\alpha}\,\ell',\quad \Omega=\sqrt{2\,M\,a+\frac{M^2\,\omega^2_{c}}{\alpha^2}},\quad j=\sqrt{\frac{(\ell-\Phi)^2}{\alpha^2}+2\,M\,b}.
\label{9}
\end{eqnarray}

Equation (\ref{8}) is a second-order differential equation, and there are several methods available for solving it. In this context, our main emphasis is on the parametric NU-method. This method has proven to be effective in solving quantum mechanical problems in the existing literature \cite{bb24,bb25,bb26}.

Thus, to solve the Eq. (\ref{8}) using the parametric NU-method, we introduce a new variables via $s=\Omega\,r^2$ into the equation (\ref{8}), we obtain the following second-order differential equation:
\begin{equation}
\psi''(s)+\frac{1}{s}\,\psi'(s)+\frac{1}{s^2}\,\Big(-\xi_1\,s^2+\xi_2\,s-\xi_3\Big)\,\psi(s)=0,
\label{10}
\end{equation}
where we have set the different parameters as
\begin{equation}
\xi_1=\frac{1}{4}\quad,\quad \xi_2=\frac{\Lambda}{4\,\Omega}\quad,\quad \xi_3=\frac{j^2}{4}.
\label{11}
\end{equation}

Thereby, following the procedure in \cite{bb24,bb25,bb26}), we obtain the following expression of the energy eigenvalue associated with the modes $\{n, \ell\}$ given by 
\begin{eqnarray}
E_{n,\ell}=c+\frac{\omega_c}{\alpha^2}\,|\ell-\Phi|+\sqrt{\frac{2\,a}{M}+\frac{\omega^2_{c}}{\alpha^2}}\,\Bigg[2\,n+\sqrt{\frac{(\ell-\Phi)^2}{\alpha^2}+2\,M\,b}+1\Bigg],
\label{13}
\end{eqnarray}
where $n=0,1,2,...$.

\begin{figure}
\begin{subfigure}[b]{0.45\textwidth}
\includegraphics[width=2.8in,height=1.3in]{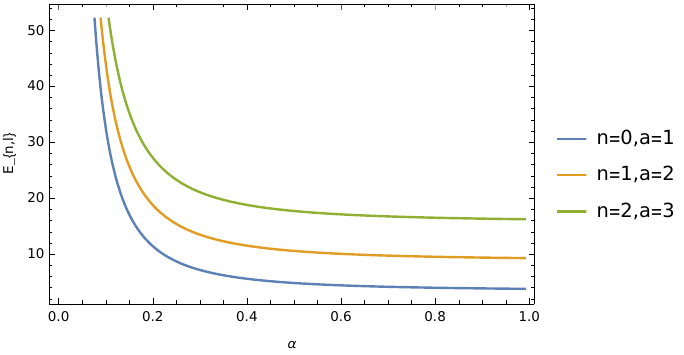}
\caption{$c=0$, $l=1=M=b=B=|e|$, $\Phi=3/4$}
\label{fig: 2(a)}
\end{subfigure}
\hfill
\begin{subfigure}[b]{0.45\textwidth}
\includegraphics[width=2.8in,height=1.3in]{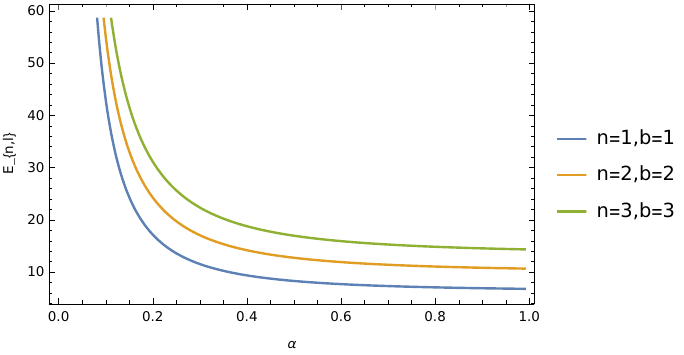}
\caption{$c=0$, $l=1=M=a=B=|e|$, $\Phi=3/4$}
\label{fig: 2(b)}
\end{subfigure}
\hfill\\
\begin{subfigure}[b]{0.45\textwidth}
\includegraphics[width=2.8in,height=1.3in]{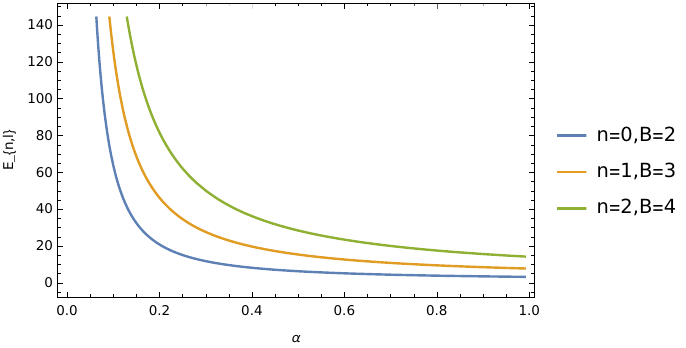}
\caption{$c=0$, $l=1=M=a=b=|e|$, $\Phi=3/4$}
\label{fig: 2(c)}
\end{subfigure}
\hfill
\begin{subfigure}[b]{0.45\textwidth}
\includegraphics[width=2.8in,height=1.3in]{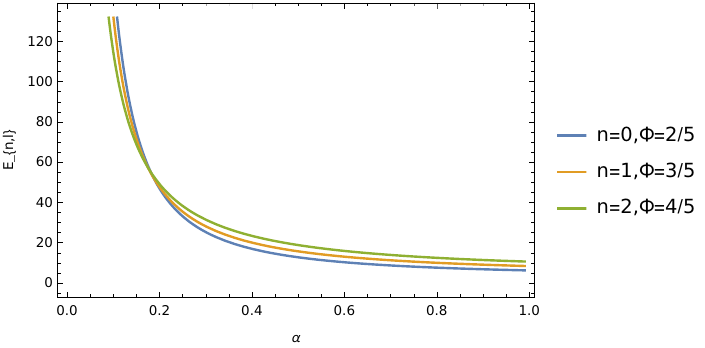}
\caption{$c=0$, $l=1=M=a=b=|e|$, $B=2$}
\label{fig: 2(d)}
\end{subfigure}
\caption{Energy levels with topological defect parameter $\alpha$ for different values of potential parameters $a,b$, magnetic field $B$, and quantum flux $\Phi$.}
\label{fig:2}
\end{figure}

The normalized radial wave functions are given by
\begin{equation}
\psi_{n,\ell} (s)=D_{n,\ell}\,s^{\frac{j}{2}}\,e^{-\frac{s}{2}}\,L^{(j)}_{n}(s).
\label{14}
\end{equation}
In terms of $r$, the radial wave function will be
\begin{equation}
\psi_{n,\ell} (r)=D_{n,\ell}\,(\Omega)^{\frac{j}{2}}\,r^{j}\,e^{-\frac{1}{2}\,\Omega\,r^2}\,L^{(j)}_{n} (\Omega\,r^2),
\label{15}
\end{equation}
where $\Omega$ and $j$ are given in equation (\ref{9}), and $D_{n,\ell}$ is the normalization constant given by
\begin{equation}
D_{n,\ell}=\sqrt{\frac{2\,\Omega}{\alpha}}\,\sqrt{\frac{n!}{(n+j)!}}.
\label{17}
\end{equation}

\begin{figure}
\begin{subfigure}[b]{0.45\textwidth}
\includegraphics[width=2.8in,height=1.3in]{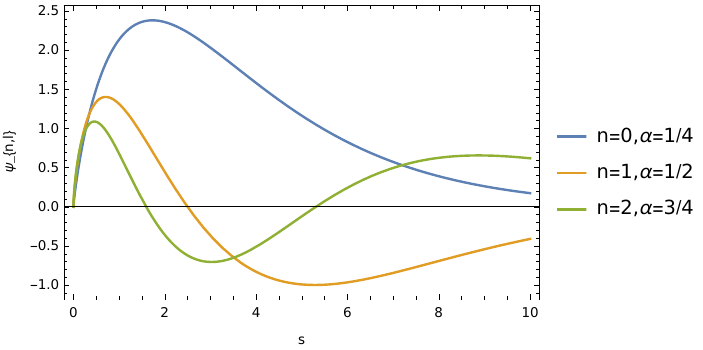}
\caption{$\ell=1=M=a=b=B=|e|$, $\Phi=3/4$}
\label{fig: 3 (a)}
\end{subfigure}
\hfill
\begin{subfigure}[b]{0.45\textwidth}
\includegraphics[width=2.7in,height=1.3in]{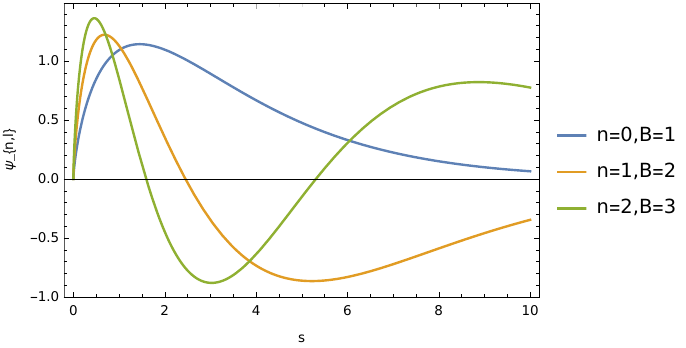}
\caption{$\ell=1=M=a=b=|e|$, $\Phi=3/4=\alpha$}
\label{fig: 3 (b)}
\end{subfigure}
\hfill\\
\begin{subfigure}[b]{0.45\textwidth}
\includegraphics[width=2.7in,height=1.3in]{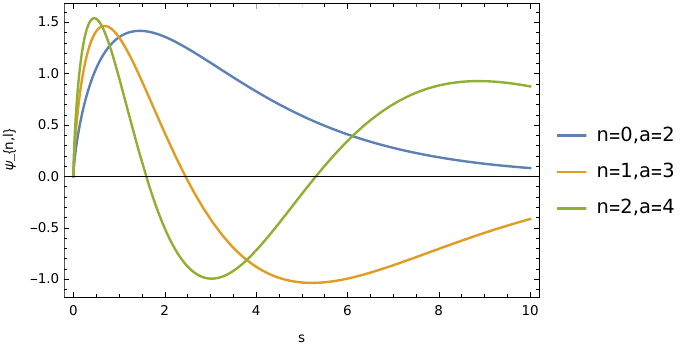}
\caption{$\ell=1=M=b=|e|$, $\Phi=3/4=\alpha$, $B=2$}
\label{fig: 3 (c)}
\end{subfigure}
\hfill
\begin{subfigure}[b]{0.45\textwidth}
\includegraphics[width=2.7in,height=1.3in]{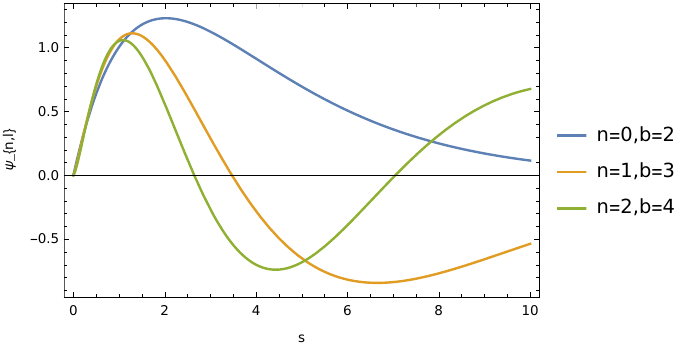}
\caption{$\ell=1=M=B=|e|$, $a=2$, $\Phi=3/4=\alpha$}
\label{fig: 3 (d)}
\end{subfigure}
\caption{Radial wave function $\psi_{n,\ell}$ with $s$ for different values of topological defect parameter $\alpha$, potential parameters $a,b$, and the magnetic field $B$.}
\label{fig:3}
\end{figure}

Equation (\ref{13}) describes the energy eigenvalues for non-relativistic quantum charged particles in the presence of a cyclotron frequency $\omega_{c}$. Meanwhile, equations (\ref{14}) through (\ref{17}) represent the normalized radial wave functions associated with these particles. It is evident that the energy levels and wave functions are influenced by topological defects present in the cosmic string space-time, denoted by the parameter $\alpha$, as well as the presence of an anharmonic oscillator potential. The presence of a topological defect, characterized by $\alpha$, leads to shifts in the energy levels and disrupts the previously existing degeneracy. We have depicted graphical representations of these energy levels in figure \ref{fig:2} and the normalized radial wave functions in figure \ref{fig:3}, with variations in different parameters. In the limit as $\alpha \to 1$, Equations (\ref{13}) through (\ref{14}) simplify to the Landau solution when the anharmonic oscillator potential is considered.

\subsection{\bf Application to Molecular potential Models}
\label{sec2.1}

In this section, we will utilize the eigenvalue solution derived earlier to develop solutions for specific types of physical potential models. These models have wide-ranging applications in addressing practical problems and are crucial for understanding various physical phenomena.

\vspace{0.1cm}
{\bf Case A: Harmonic Oscillator Potential}
\vspace{0.1cm}

The harmonic oscillator potential can be recovered from the potential (\ref{9}) by setting the potential parameters $b=0$, $c=0$, and $a=\frac{1}{2}\,M\,\omega^2$, one will have $V (r)=\frac{1}{2}\,M\,\omega^2\,r^2$.

Hence, by incorporating the harmonic oscillator potential into the quantum system and subsequently solving the wave equation while considering other relevant physical parameters, the resulting energy eigenvalue expression for the harmonic oscillator is as follows:
\begin{eqnarray}
E_{n,\ell}=\frac{\omega_c}{\alpha^2}|\ell-\Phi|+\sqrt{\omega^2+\frac{\omega^2_{c}}{\alpha^2}}\Big(2\,n+\frac{|\ell-\Phi|}{\alpha}+1).
\label{aa1}
\end{eqnarray}
And the radial wave function will be
\begin{eqnarray}
\psi_{n,\ell} (r)=\sqrt{\frac{2}{\alpha}}\,\sqrt{\frac{n!}{(n+\frac{|\ell-\Phi|}{\alpha})!}}\,(\Omega)^{\frac{1}{2}+\frac{|\ell-\Phi|}{2\,\alpha}}\,r^{\frac{|\ell-\Phi|}{\alpha}}\,e^{-\frac{1}{2}\Omega r^2}\,L^{(\frac{|\ell-\Phi|}{\alpha})}_{n}(\Omega r^2),\quad \Omega=\sqrt{\omega^2+\frac{\omega^2_{c}}{\alpha^2}}.
\label{aa2}
\end{eqnarray}
In the limit $\alpha \to 1$, Eqs.(\ref{aa1}) and (\ref{aa2}) gives the Landau solutions of a harmonic oscillator in the presence of magnetic and quantum flux fields.

\vspace{0.1cm}
{\bf Case B: Pseudoharmonic Potential}
\vspace{0.1cm} 

The pseudoharmonic potential can be recovered from the potential (\ref{6}) by setting the parameters $a=\frac{D_e}{r^{2}_{0}}$, $b=D_e\,r^{2}_{0}$, and $c=-2\,D_e$. Thus, we obtain the following form of the potential \cite{SMI,RSS,FC2,AC,SMI5} given by
\begin{equation}
V(r)=\frac{D_e}{r^{2}_{0}}\,r^2+\frac{D_e\,r^{2}_{0}}{r^2}-2\,D_e=D_e\,\Bigg(\frac{r}{r_0}-\frac{r_0}{r}\Bigg)^2.
\label{bb1}
\end{equation}
Here $D_e$ is the dissociation energy between atoms in a solid and $r_0$ is the equilibrium inter-nuclear separation. This potential is of great importance not only in physics but also in chemistry and is useful to describe the interactions of some diatomic molecules \cite{AC,SMI5}.

Therefore, using the above potential into the quantum and after solving the wave equation, one can find the energy eigenvalues given by
\begin{eqnarray}
E_{n,\ell}=-2\,D_e+\frac{\omega_c}{\alpha^2}\,|\ell-\Phi|+\sqrt{\frac{2\,D_e}{M\,r^2_{0}}+\frac{\omega^2_{c}}{\alpha^2}}\Big(2\,n+\sqrt{\frac{(\ell-\Phi)^2}{\alpha^2}+2\,M\,D_e\,r^2_{0}}+1\Big).
\label{bb2}
\end{eqnarray}
And that the radial wave function will be
\begin{eqnarray}
\psi_{n,\ell} (r)=\sqrt{\frac{2\,\Omega\,(n!)}{\alpha\,(n+j)!}}\,s^{\frac{j}{2}}\,e^{-\frac{s}{2}}\,L^{(j)}_{n} (s),\quad 
\Omega=\sqrt{\frac{2\,M\,D_e}{r^2_{0}}+\frac{M^2\,\omega^2_{c}}{\alpha^2}},\quad
j=\sqrt{\frac{(\ell-\Phi)^2}{\alpha^2}+2\,M\,D_e\,r^2_{0}}.
\label{bb3}
\end{eqnarray}

\vspace{0.1cm}
{\bf Case C: Shifted Pseudoharmonic Potential}
\vspace{0.1cm} 

In this case, we set the parameters $a=\frac{D_e}{r^{2}_{0}}$, $b=D_e\,r^{2}_{0}$, and $c=0$. Thus, we obtain the following potential form \cite{SMI,RSS,FC2}
\begin{eqnarray}
V(r)=\frac{D_e}{r^{2}_{0}}\,r^2+\frac{D_e\,r^{2}_{0}}{r^2}=D_e\,\Bigg(\frac{r}{r_0}-\frac{r_0}{r}\Bigg)^2+2\,D_e\quad
\label{cc1}
\end{eqnarray}
which is shifted by an amount twice the dissociation energy $D_e$ compared to pseudoharmonic potential.

Therefore, using the shifted pseudoharmonic potential into the above quantum system and after solving the wave equation, one can find the energy eigenvalues as follows:
\begin{eqnarray}
E_{n,\ell}=\frac{\omega_c}{\alpha^2}\,|\ell-\Phi|+\sqrt{\frac{2\,D_e}{M\,r^2_{0}}+\frac{\omega^2_{c}}{\alpha^2}}\Big(2\,n+\sqrt{\frac{(\ell-\Phi)^2}{\alpha^2}+2\,M\,D_e\,r^2_{0}}+1\Big).
\label{cc2}
\end{eqnarray}
The radial wave function will be same given in the Eq. (\ref{bb3}).

\vspace{0.1cm}
{\bf Case D: Inverse Square Potential}
\vspace{0.1cm}

The inverse square potential can be derived from the potential (\ref{9}) by setting the potential parameters as $a=0$ and $c=0$. This results in the expression $V(r) = \frac{b}{r^2}$, which is characterized by its repulsive behavior. Despite its singular nature, the inverse square potential holds immense significance as one of the fundamental interactions in quantum mechanics.

This potential function arises in the study of crucial problems across various domains of physics, including phenomena such as the Efimov effect \cite{hh1}, dipole-bound anions in polar molecules \cite{hh2, hh2b, hh3}, atomic interactions with charged wires \cite{hh4, hh4b}, analysis of near-horizon structures of black holes \cite{hh5, hh5b, hh5c}, and the interaction of dipoles within a cosmic string background \cite{hh6}. Moreover, numerous authors have employed the inverse square potential to model quantum systems, as evidenced by references \cite{bb18, ss8, ss9, ss10, ss11}.

Therefore, using this inverse square potential into the quantum system, we obtain the following radial equation
\begin{equation}
\psi''(r)+\frac{1}{r}\,\psi'(r)+\Bigg(\Lambda-\frac{j^2}{r^2}-\frac{M^2\,\omega^2_{c}}{\alpha^2}\,r^2\Bigg)\,\psi(r)=0,
\label{ss1}
\end{equation}
where $\Lambda, j$ are given in equation (\ref{9}).

Following the previous procedure, one can obtain the expression of the energy eigenvalue given by
\begin{eqnarray}
E_{n,\ell}=\frac{\omega_c}{\alpha}\Bigg[2\,n+1+\sqrt{\frac{(\ell-\Phi)^2}{\alpha^2}+2\,M\,b}+\frac{|\ell-\Phi|}{\alpha}\Bigg],\quad
\label{ss2}
\end{eqnarray}
where $n=0,1,2,...$.

The normalized radial wave functions are given by
\begin{equation}
\psi_{n,\ell} (x)=\sqrt{\frac{2\,M\,\omega_c}{\alpha^2}}\,\sqrt{\frac{n!}{(n+j)!}}\,x^{\frac{j}{2}}\,e^{-\frac{x}{2}}\,L^{(j)}_{n} (x),
\label{ss3}
\end{equation}
where $x=\Big(\frac{M\,\omega_c}{\alpha}\Big)\,r^2$.

In the limit $\alpha \to 1$, the space-time geometry under consideration becomes Minkowski flat space and the eigenvalue solution reduces to the Landau levels in the presence of external fields with this inverse square potential. 

\section{Thermo-magnetic Properties of the Quantum System}
\label{sec3}

In this section, our focus on investigating the thermodynamic and magnetic properties of the quantum system under investigation. We calculate different physical quantities in thermodynamics and magnetic system and analyze the effects of disclinations, potential, and the external magnetic and quantum flux fields.

\subsection{\bf Thermodynamic properties}
\label{sec3.1}

Here, we compute the partition function $Z(\beta)$ for this system at a finite temperature $T \neq 0$, utilizing the energy expression $E_{n,\ell}$ as given in Eq. (\ref{13}). Subsequently, we will delve into the analysis of various other thermodynamic properties, including the vibrational free energy, mean energy, specific heat capacity, and entropy.

The energy eigenvalue (\ref{13}) keeping the orbital quantum number $\ell$ to be fixed can express as (setting $c=0$)
\begin{equation}
E_{n}=P+2\,\omega_0\,n,
\label{dd1}
\end{equation}
where
\begin{eqnarray}
P=Q+\omega_0\,\Bigg(\sqrt{\frac{(\ell-\Phi)^2}{\alpha^2}+2\,M\,b}+1\Bigg),\quad 
\omega_0=\sqrt{\frac{2\,a}{M}+\frac{\omega^2_{c}}{\alpha^2}},\quad Q=\frac{\omega_c}{\alpha^2}\,|\ell-\Phi|.
\label{dd}
\end{eqnarray}

\vspace{0.1cm}
{\bf Partition Function}
\vspace{0.1cm}

The partition function can be obtained from the following relation \cite{COE,COE2,ANI,CSJ,SR}
\begin{equation}
Z(\beta)=\sum^{n=\infty}_{n=0}\,e^{-\beta\,E_{n}}.
\label{dd2}
\end{equation}

\begin{figure}
\begin{subfigure}[b]{0.45\textwidth}
\includegraphics[width=2.8in,height=1.3in]{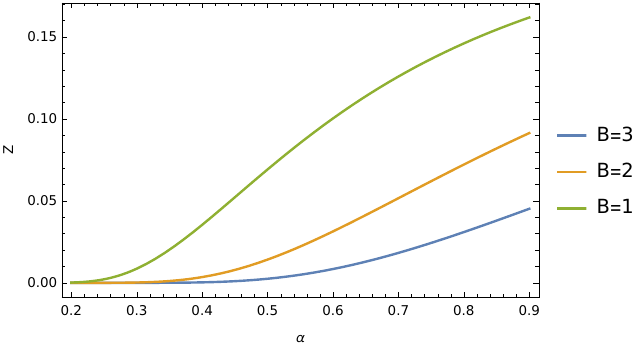}
\caption{$\ell=1=M=a=b=|e|$, $\Phi=1/2=\beta$}
\label{fig: 4 (a)}
\end{subfigure}
\hfill
\begin{subfigure}[b]{0.45\textwidth}
\includegraphics[width=2.8in,height=1.3in]{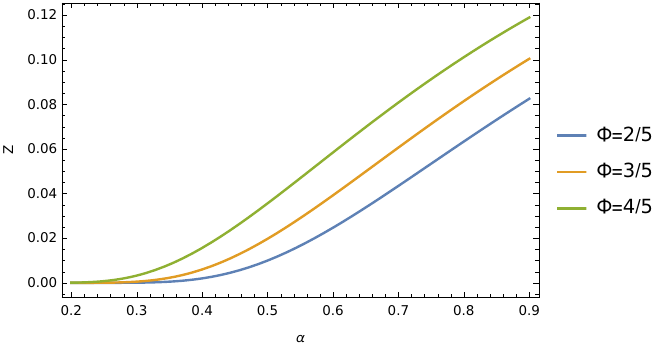}
\caption{$\ell=1=M=a=b=|e|$, $B=2$, $\beta=0.5$}
\label{fig: 4 (b)}
\end{subfigure}
\caption{Partition function $Z$ with topological defect parameter $\alpha$ for different values of the magnetic field $B$ and quantum flux $\Phi$.}
\label{fig:4}
\begin{subfigure}[b]{0.45\textwidth}
\includegraphics[width=2.8in,height=1.3in]{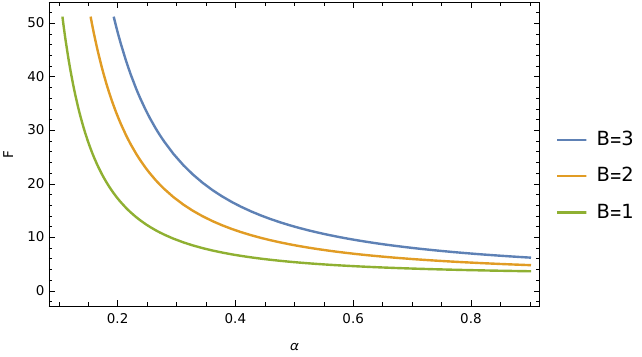}
\caption{$\ell=1=M=a=b=|e|$, $\Phi=1/2=\beta$}
\label{fig: 5 (a)}
\end{subfigure}
\hfill
\begin{subfigure}[b]{0.45\textwidth}
\includegraphics[width=2.8in,height=1.3in]{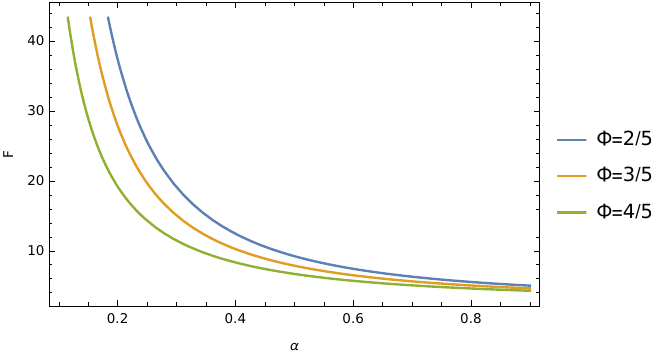}
\caption{$\ell=1=M=a=b=|e|$, $B=2$, $\beta=0.5$}
\label{fig: 5 (b)}
\end{subfigure}
\caption{Vibrational free energy with topological defect parameter $\alpha$ for different values of the magnetic field $B$ and quantum flux $\Phi$.}
\label{fig:5}
\end{figure}

Using the expression (\ref{dd1}), we have
\begin{equation}
Z=\frac{e^{-\beta\,(\omega_0\,j+Q)}}{2\,\sinh (\beta\,\omega_0)}\quad ,\quad j=\sqrt{\frac{(\ell-\Phi)^2}{\alpha^2}+2\,M\,b}
\label{dd3}
\end{equation}
which means that the partition function $Z$ depends on the potential strengths $a, b$, the topological defects of cosmic string characterized by the parameter $\alpha$, and the external magnetic $B$ and quantum flux fields $\Phi_{AB}$. It also depends on the quantum number $\ell$ and the absolute temperature $T$, and changes with them. The behavior of the partition function is demonstrated in Fig.\ref{fig:4}.

\vspace{0.1cm}
{\bf Vibrational Free Energy}
\vspace{0.1cm}

%It is defined in terms of partition function $Z(\beta)$ as
%\begin{equation}
%F(\beta)=-\frac{1}{\beta}\,\mbox{ln} Z(\beta).
%\label{dd4}
%\end{equation}

Using the partition function expression (\ref{dd3}), we obtain the vibrational free energy given by 
\begin{eqnarray}
F=-\frac{1}{\beta}\,\mbox{ln} Z=\frac{\omega_c}{\alpha^2}\,|\ell-\Phi|+\sqrt{\frac{2\,a}{M}+\frac{\omega^2_{c}}{\alpha^2}}\,\sqrt{\frac{(\ell-\Phi)^2}{\alpha^2}+2\,M\,b}+\frac{1}{\beta}\,\mbox{ln}\,\Bigg[2\,\sinh \Big(\beta\,\sqrt{\frac{2\,a}{M}+\frac{\omega^2_{c}}{\alpha^2}}\Big)\Bigg].
\label{dd5}
\end{eqnarray}
which depends on the topological defect of cosmic string characterized by the parameter $\alpha$, the potential parameters $a, b$, and the external magnetic $B$ and quantum flux fields $\Phi_{AB}$. Furthermore, it depends on the quantum number $\ell$ and changes with the absolute temperature $T$. 
In Fig.\ref{fig:5} the behavior of vibrational free energy is plotted.

\vspace{0.1cm}
{\bf Mean Energy}
\vspace{0.1cm}

%It is defined in terms of partition function $Z(\beta)$ as
%\begin{equation}
%U(\beta)=-\frac{\partial\,\mbox{ln}\,Z(\beta)}{\partial\,\beta}
%\label{dd6}
%\end{equation}

\begin{figure}
\begin{subfigure}[b]{0.45\textwidth}
\includegraphics[width=2.8in,height=1.3in]{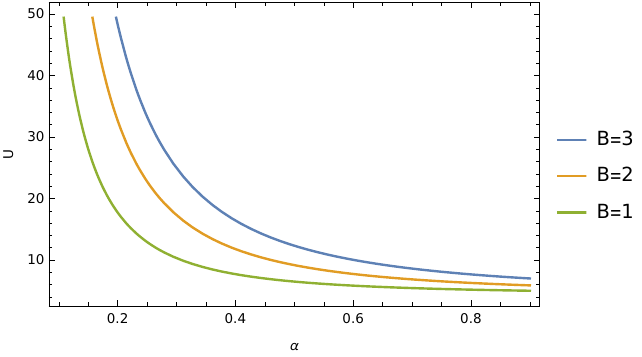}
\caption{$\ell=1=M=a=b=B=|e|$, $\Phi=1/2=\beta$}
\label{fig: 6 (a)}
\end{subfigure}
\hfill
\begin{subfigure}[b]{0.45\textwidth}
\includegraphics[width=2.8in,height=1.3in]{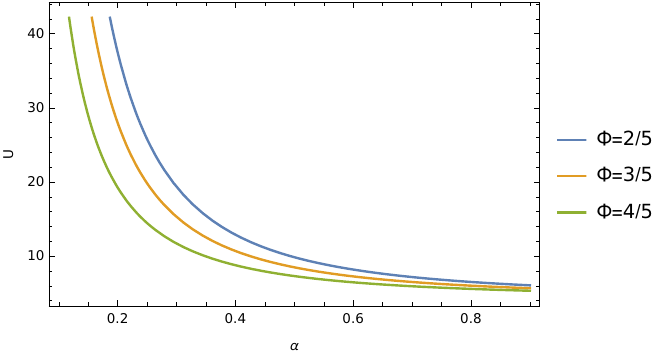}
\caption{$\ell=1=M=a=b=|e|$, $B=2$, $\beta=0.5$}
\label{fig: 6 (b)}
\end{subfigure}
\caption{Mean free energy with topological defect parameter $\alpha$ for different values of the magnetic field $B$ and quantum flux $\Phi$.}
\label{fig:6}
\begin{subfigure}[b]{0.45\textwidth}
\includegraphics[width=2.8in,height=1.3in]{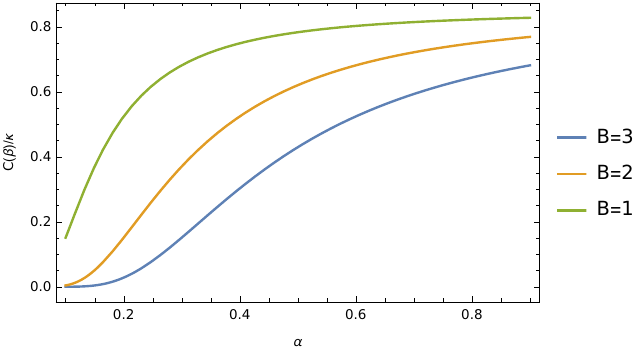}
\caption{$M=a=|e|$, $\beta=0.5$}
\label{fig: 7 (a)}
\end{subfigure}
\hfill
\begin{subfigure}[b]{0.45\textwidth}
\includegraphics[width=2.8in,height=1.3in]{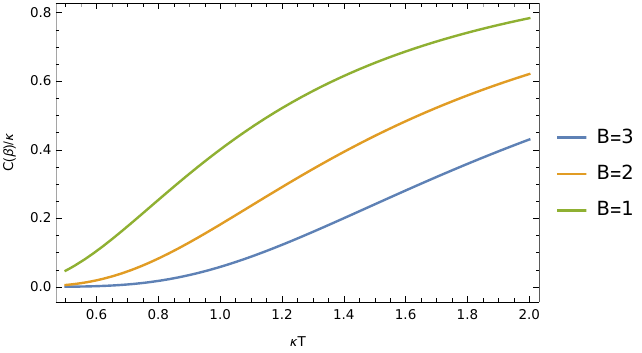}
\caption{$M=a=|e|$, $\alpha=1/2$}
\label{fig: 7 (b)}
\end{subfigure}
\caption{Specific heat capacity with topological defect parameter $\alpha$ and absolute temperature $T$ for different values of the magnetic field $B$ and quantum flux $\Phi$.}
\label{fig:7}
\end{figure}

Using the partition function expression (\ref{dd3}), we obtain the mean energy given by 
\begin{eqnarray}
U=-\frac{\partial\,\mbox{ln} Z}{\partial \beta}=\frac{\omega_c}{\alpha^2}\,|\ell-\Phi|+\sqrt{\frac{2\,a}{M}+\frac{\omega^2_{c}}{\alpha^2}}\,\Bigg[\sqrt{\frac{(\ell-\Phi)^2}{\alpha^2}+2\,M\,b}+\coth \Big(\beta\,\sqrt{\frac{2\,a}{M}+\frac{\omega^2_{c}}{\alpha^2}}\Big)\Bigg].
\label{dd7}
\end{eqnarray}
which depends on the topological defect of cosmic string characterized by the parameter $\alpha$, the potential parameters $a, b$, and the external magnetic $B$ and quantum flux $\Phi_{AB}$ fields. It changes with the quantum number $\ell$ and the absolute temperature $T$. In Fig.\ref{fig:6} the behavior of the mean energy is demonstrated.

\vspace{0.1cm}
{\bf Specific Heat Capacity}
\vspace{0.1cm}

%It is defined in terms of partition function $Z(\beta)$ as
%\begin{equation}
%C(\beta)=\kappa\,\beta^2\,\frac{\partial^2\,\mbox{ln}\,Z(\beta)}{\partial\,\beta^2}.
%\label{dd8}
%\end{equation}

Using the partition function expression (\ref{dd3}), we obtain the specific heat capacity given by
\begin{equation}
\frac{C}{\kappa}=\beta^2\,\Bigg(\frac{\partial^2\,\mbox{ln}\,Z}{\partial\beta^2}\Bigg)=\frac{\beta^2\,\Big(\frac{2\,a}{M}+\frac{\omega^2_{c}}{\alpha^2}\Big)}{\Bigg[\sinh \Big(\beta\,\sqrt{\frac{2\,a}{M}+\frac{\omega^2_{c}}{\alpha^2}}\Big)\Bigg]^2}
\label{dd9}
\end{equation}
which depends on the topological defect of cosmic string characterized by the parameter $\alpha$, the potential parameter $a$, the external magnetic field $B$, and the absolute temperature $T$. The specific heat capacity is plotted in figure \ref{fig:7}.

\vspace{0.1cm}
{\bf Entropy }
\vspace{0.1cm}

%It is defined in terms of partition function $Z(\beta)$ as
%\begin{equation}
%S(\beta)=\kappa\,\mbox{ln}\,Z(\beta)-\beta\,\kappa\,\frac{\partial\,\mbox{ln}\,Z(\beta)}{\partial\,\beta}.
%\label{dd10}
%\end{equation}

Using the partition function expression (\ref{dd3}), we obtain the specific heat capacity given by
\begin{eqnarray}
\frac{S}{\kappa}=\mbox{ln} Z-\beta\,\Bigg(\frac{\partial\,\mbox{ln} Z}{\partial\beta}\Bigg)=\beta\,\sqrt{\frac{2\,a}{M}+\frac{\omega^2_{c}}{\alpha^2}}\,\coth \Big(\beta\,\sqrt{\frac{2\,a}{M}+\frac{\omega^2_{c}}{\alpha^2}}\Big)
-\mbox{ln}\,\Bigg[2\,\sinh \Big(\beta\,\sqrt{\frac{2\,a}{M}+\frac{\omega^2_{c}}{\alpha^2}}\Big)\Bigg].
\label{dd11}
\end{eqnarray}
which depends on the topological defect of cosmic string characterized by the parameter $\alpha$, the potential parameter $a$, the external magnetic field $B$, the quantum number $l$, and the absolute temperature $T$. 

%\section{Magnetic Properties of the Quantum System: Persistent Currents, Magnetization and Magnetic Susceptibility }
%\label{sec5}

%In this section, we will explore the magnetic properties of the quantum system under examination. Our analysis will encompass the calculation of persistent currents, magnetization, and magnetic susceptibility, both at absolute zero temperature ($T=0$) and at finite temperature ($T\neq 0$). It's important to note that we have introduced external magnetic and quantum flux fields through a minimal substitution in the wave equation, and our earlier findings have revealed that the eigenvalue solutions are affected by these fields, in addition to the linear topological defect of cosmic strings.

\subsection{\bf Magnetic properties}
\label{sec3.2}

Now, we study the magnetic properties of the quantum system under examination. Our analysis will encompass the calculation of persistent currents, magnetization, and magnetic susceptibility, both at absolute zero temperature ($T=0$) and at finite temperature ($T\neq 0$).  

\vspace{0.1cm}
{\bf Persistent Currents}
\vspace{0.1cm}

It is well-known in condensed matter physics \cite{aa57,aa58,aa59} that the dependence of the energy eigenvalue on the geometric quantum phase give rise to a persistent currents. The expression for the total persistent currents at zero temperature, $T=0$ is given by \cite{aa60,SR,fa7,fa8}
\begin{equation}
I=\sum_{n,\ell}\,I_{n,\ell}\quad \mbox{and} \quad I_{n,\ell}=-\frac{\partial E_{n,\ell}}{\partial \Phi_{AB}}
\label{dd12}
\end{equation}
is called the Byers-Yang relation \cite{aa57}.

Using the expression Eq. (\ref{13}), we obtain the Byers-Yang relation by
\begin{equation}
I_{n,\ell}=\frac{e\,\omega_c}{2\,\pi\,\alpha^2}+\frac{|e|\,|\ell-\Phi|}{2\,\pi\,\alpha^2}\,\frac{\sqrt{\frac{2\,a}{M}+\frac{\omega^2_{c}}{\alpha^2}}}{\sqrt{\frac{(\ell-\Phi)^2}{\alpha^2}+2\,b\,M}}.
\label{dd13}
\end{equation}

The persistent currents at finite temperature $T \neq 0$ can be calculated using the following relation \cite{SR,COE2}
\begin{equation}
I_{n,\ell} (\beta)=-\frac{\partial F}{\partial \Phi_{AB}}.
\label{ee1}
\end{equation}
Using the free energy expression (\ref{dd5}), one will get the same expression given in Eq. (\ref{dd13}). Thus, the persistent currents in the present analyzed quantum system is independent of the temperature.

\vspace{0.1cm}
{\bf Magnetization}
\vspace{0.1cm}

The magnetization of a quantum system at zero temperature $T=0$ in a quantum state $\{n, l\}$ is defined by \cite{fa7,fa8}
\begin{equation}
M_{n,\ell} (B, \Phi_{AB})=-\frac{\partial E_{n,\ell}}{\partial B}
\label{dd14}
\end{equation}

Using the expression Eq. (\ref{13}), we obtain the magnetization as follows   
\begin{eqnarray}
M_{n,\ell}=-\frac{|e|\,|\ell-\Phi|}{2\,M\,\alpha^2}-\frac{e^2\,B}{4\,M^2\,\alpha^2}\frac{\Big(2\,n+1+\sqrt{\frac{(\ell-\Phi)^2}{\alpha^2}+2\,b\,M}\Big)}{\sqrt{\frac{2\,a}{M}+\frac{\omega^2_{c}}{\alpha^2}}}.
\label{dd15}
\end{eqnarray}

At finite temperature $T \neq 0$, the magnetization of a quantum system is defined by \cite{SR,COE2}
\begin{equation}
M=\frac{1}{\beta\,Z}\,\frac{\partial Z}{\partial B}=-\frac{\partial F}{\partial B}.
\label{ee2}
\end{equation}

Using the free energy expression (\ref{dd5}), one can obtain
\begin{eqnarray}
M=-\frac{e^2\,B}{4\,M^2\,\alpha^2\,\sqrt{\frac{2\,a}{M}+\frac{\omega^2_{c}}{\alpha^2}}}\,\Bigg[\sqrt{\frac{(\ell-\Phi)^2}{\alpha^2}+2\,b\,M}+\coth \Big(\beta\,\sqrt{\frac{2\,a}{M}+\frac{\omega^2_{c}}{\alpha^2}}\Big)\Bigg]-\frac{|e|\,|\ell-\Phi|}{2\,M\,\alpha^2}.
\label{ee3}
\end{eqnarray}

\vspace{0.1cm}
{\bf Magnetic Susceptibility }
\vspace{0.1cm}

The magnetic susceptibility at zero temperature $T=0$ in a quantum state $\{n, l\}$ is defined by \cite{fa7,fa8}
\begin{equation}
\chi_{n,\ell} (B, \Phi_{AB})=\frac{\partial M_{n,\ell}}{\partial B}.
\label{dd16}
\end{equation}

Using the expression Eq. (\ref{dd15}), we obtain the magnetic susceptibility as follows  
\begin{equation}
\chi_{n,\ell}=-\frac{2\,a\,e^2}{4\,M^3\,\alpha^2}\,\frac{\Big(2\,n+1+\sqrt{\frac{(\ell-\Phi)^2}{\alpha^2}+2\,b\,M}\Big)}{\Big(\frac{2\,a}{M}+\frac{\omega^2_{c}}{\alpha^2}\Big)^{3/2}}
\label{dd17}
\end{equation}

At finite temperature $T \neq 0$, the magnetic susceptibility is defined by \cite{SR,COE2}
\begin{equation}
\chi=\frac{\partial M}{\partial B}
\label{ee4}
\end{equation}
that can be obtained using the expression (\ref{ee3}).

It becomes evident that the magnetic characteristics of the quantum system, including persistent currents, magnetization, and magnetic susceptibility, are influenced by the topological defects of the cosmic string. These effects are governed by the parameter $\alpha$ and the potential strengths $(a, b)$ associated with the anharmonic oscillator potential. Additionally, the presence of external magnetic and quantum fields $(B, \Phi_{AB})$ plays a significant role in modifying these properties, both at zero temperature and in the presence of finite temperature conditions. An intriguing observation is that persistent currents within the quantum system remain unaffected by variations in temperature. Conversely, other magnetic properties exhibit changes in response to temperature variations.

\section{Entropic Information}
\label{sec4}

In this section, we will analyze the entropic information measures for the cases presented in the previous sections. Information theories have played an important role in the development of modern technology and the advancement of technologies in current quantum computing. Among these theories, Shannon entropy is the one that has been most studied, due to its simplicity and its significant results in investigating the location of particles in different quantum systems \cite{Wang:2023xhy,Santana-Carrillo:2022gmu,Gil-Barrera:2022ayj,Lima:2022dts,Lima1,Moreira:2023fui,Benabdallah:2023gku,Lopez-Garcia:2023dah, Mondal:2023wmf}. The Shannon entropy for position space is 
\begin{equation}
S^n_{\textbf{r}}=-\int\vert\Psi_{n}(\textbf{r})\vert^{2}\ln\vert\Psi_{n}(\textbf{r})\vert^{2}d^D\textbf{r},\label{16}
\end{equation} 
where $D$ are the dimensions that affect the system's information measurements. Through the Fourier transform of the system's probability density
\begin{align}
    \vert\Psi_n(\textbf{p})\vert^2=\frac{1}{(2\pi)^{D/2}}\int\,\vert\Psi(\textbf{r})\vert^2\text{e}^{-i\textbf{p}\cdot\textbf{r}}\,d^D\textbf{r},
\end{align}
we can define Shannon entropy for momentum space in the form
\begin{equation}
S^n_{\textbf{p}}=-\int\vert\Psi_{n}(\textbf{p})\vert^{2}\ln\vert\Psi_{n}(\textbf{p})\vert^{2}d^D\textbf{p}.
\end{equation}
Furthermore, Shannon's information entropies generate an uncertainty relationship that must be obeyed by the system
\begin{equation}
S^n_{\textbf{r}}+S^n_{\textbf{p}}\geq D(1+\text{ln}\pi).
\end{equation}
This uncertainty relationship was first proposed by Beckner, Bialynicki-Birula, and Mycielski, and is also known as BBM uncertainty \cite{Beckner,Bialy}. In our model, the dimension $\phi$ is cyclical and does not interfere with the information measurements. Therefore, the only dimension that interests us is that of the radial direction, $r$  ($D=1$).

\vspace{0.1cm}
{\bf Case A}
\vspace{0.1cm}

For the case of the harmonic oscillator potential, Shannon entropy is analyzed through Tab.\ref{tab1}. When we increase the value of $\omega$ (angular frequency of oscillation of the harmonic oscillator) the information entropy in the position space tends to decrease. This indicates that the chance of locating the particle increases as $\omega$ increases. On the other hand, the information entropy decreases for the momentum space, i.e., the uncertainty of the particle's location will be greater. Furthermore, the sum of the two entropies ($S^n_{r}+S^n_{p}$) increases, which indicates a process of delocalization of the particle as $\omega$ increases.

When we increase the value of the magnetic field $B$ something interesting happens, the information entropy decreases until it becomes negative. This comes from the fact that the probability density is extremely localized, which leads to less uncertainty in the location of the particle. The entropy for the momentum space behaves contrary to that of the position, indicating less certainty in the location of the particle in the momentum space. Total entropy increases, increasing uncertainty in particle location (delocalization). Something similar happens when we decrease the value of $\alpha$ (topological defect), i.e., $\alpha\rightarrow0$. In other words, when we reduce the influence of the topological defect, the uncertainty in the location of the particle will be greater.

Finally, when we increase the influence of the quantum flow $\Phi$, the entropic information increases in the position space. The opposite happens in momentum space, where entropy decreases. This indicates an increase in the probability of locating the particle in momentum space and a decrease in position space. The sum of the two entropies (general entropy of the system) tends to decrease as we increase the value of the quantum flux, i.e., the greater the certainty in the location of the particle.

Shannon entropy increases when we analyze higher energy states. For a more direct analysis, we only consider the three lowest energy levels. Note that the higher the energy of the system, the greater the uncertainties in the location of the particle. At all energy levels, the total entropy of the system obeys the BBM uncertainty relation.

\begin{table}[ht!]
\centering
\begin{tabular}{|c|c|c|c|c|c|c|c|c|}\hline
\hline
$n$ & $\alpha$ & $\omega$ & $B$ & $\Phi$ & $S_{r}$ & $S_{p}$ & $S_{r}+S_{p}$ & $1+\ln\pi$\\ \hline
\hline
  & 3/4 & 1 & 1 &  3/4  & 0.39417  & 2.18524  & 2.57941 & \\
0 & 3/4 & 2 & 1 &  3/4  & 0.21107  & 2.62431  & 2.83539 & 2.14473 \\
  & 3/4 & 3 & 1 &  3/4  & 0.05520  & 3.06695  & 3.12216 & \\ \hline
  & 3/4 & 1 & 1 &  3/4  & 0.39417  & 2.18524  & 2.57941 & \\
0 & 3/4 & 1 & 2 &  3/4  & 0.12627  & 2.85657  & 2.98284 & 2.14473 \\
  & 3/4 & 1 & 3 &  3/4  & -0.05872 & 3.43706  & 3.37834 & \\ \hline
  & 3/4 & 1 & 1 &  3/4  & 0.39417  & 2.18524  & 2.57941 & \\
0 & 1/2 & 1 & 1 &  3/4  & 0.27031  & 2.30919  & 2.57952 & 2.14473 \\
  & 1/4 & 1 & 1 &  3/4  & -0.00994 & 2.88589  & 2.87595 & \\ \hline
  & 3/4 & 1 & 1 &  3/4  & 0.39417  & 2.18524  & 2.57941 & \\
0 & 3/4 & 1 & 1 &  1/2  & 0.36449  & 2.60346  & 2.96796 & 2.14473 \\
  & 3/4 & 1 & 1 &  1/4  & 0.30282  & 3.52935  & 3.83218 & \\ \hline
  \hline
  & 3/4 & 1 & 1 &  3/4  & 0.58994  & 2.49844  & 3.08840 & \\
1 & 3/4 & 2 & 1 &  3/4  & 0.40685  & 2.97947  & 3.38634 & 2.14473 \\
  & 3/4 & 3 & 1 &  3/4  & 0.25099  & 3.97897  & 4.22996 & \\ \hline
  & 3/4 & 1 & 1 &  3/4  & 0.58994  & 2.49844  & 3.08840 & \\
1 & 3/4 & 1 & 2 &  3/4  & 0.32205  & 3.70456  & 4.02662 & 2.14473 \\
  & 3/4 & 1 & 3 &  3/4  & 0.13705  & 4.46182  & 4.59889 & \\ \hline
  & 3/4 & 1 & 1 &  3/4  & 0.58994  & 2.49844  & 3.08840 & \\
1 & 1/2 & 1 & 1 &  3/4  & 0.48168  & 2.92872  & 3.41041 & 2.14473 \\
  & 1/4 & 1 & 1 &  3/4  & 0.22427  & 3.64496  & 3.86924 & \\ \hline
  & 3/4 & 1 & 1 &  3/4  & 0.58994  & 2.49844  & 3.08840 & \\
1 & 3/4 & 1 & 1 &  1/2  & 0.54448  & 3.45037  & 3.99486 & 2.14473 \\
  & 3/4 & 1 & 1 &  1/4  & 0.45891  & 5.56064  & 6.01957 & \\ \hline
    \hline
  & 3/4 & 1 & 1 &  3/4  & 0.69123  & 3.32245  & 4.01369 & \\
2 & 3/4 & 2 & 1 &  3/4  & 0.50814  & 3.97297  & 4.48112 & 2.14473 \\
  & 3/4 & 3 & 1 &  3/4  & 0.35227  & 4.64723  & 4.99951 & \\ \hline
  & 3/4 & 1 & 1 &  3/4  & 0.69123  & 3.32245  & 4.01369 & \\
2 & 3/4 & 1 & 2 &  3/4  & 0.42333  & 4.32679  & 4.75013 & 2.14473 \\
  & 3/4 & 1 & 3 &  3/4  & 0.23834  & 5.23984  & 5.47819 & \\ \hline
  & 3/4 & 1 & 1 &  3/4  & 0.69123  & 3.32245  & 4.01369 & \\
2 & 1/2 & 1 & 1 &  3/4  & 0.59233  & 3.68546  & 4.27781 & 2.14473 \\
  & 1/4 & 1 & 1 &  3/4  & 0.35013  & 4.13503  & 4.48517 & \\ \hline
  & 3/4 & 1 & 1 &  3/4  & 0.69123  & 3.32245  & 4.01369 & \\
2 & 3/4 & 1 & 1 &  1/2  & 0.63687  & 4.03671  & 4.67359 & 2.14473 \\
  & 3/4 & 1 & 1 &  1/4  & 0.53865  & 6.50506  & 7.04372 & \\ \hline
\end{tabular}\\
\caption{Numerical result of Shannon entropy for harmonic oscillator potential ($l=0$).\label{tab1}}
\end{table}

\vspace{0.1cm}
{\bf Case B and C}
\vspace{0.1cm}

For the cases of pseudoharmonic and shifted pseudoharmonic potential, the wave eigenfunctions are the same, so we can make a unique analysis of the Shannon entropy using the table \ref{tab2}. Note that when we increase the value of equilibrium inter-nuclear separation ($r_0$), the greater the information measures in the position space, i.e., the greater the uncertainties in the location of the particle. In momentum space, the smaller the information measures, i.e., the smaller the uncertainties in the location of the particle. The total entropy of the system ($S^n_{r}+S^n_{p}$) tends to decrease. This indicates that as inter-nuclear separation increases, the smaller the uncertainty in particle location will be, i.e., the more localized these particles will be.

When we increase the value of the dissociation energy ($D_e$) the smaller the entropic measurements in the position space will be, i.e., the particle will be more localized. In momentum space, the entropic measurements increase, i.e., less localized. The total entropy of the system tends to increase, indicating greater uncertainty in the location of the particle. The same happens when we increase the value of the magnetic field $B$.

When we reduce the influence of the quantum flow $\Phi$ and the topological defect $\alpha$, the Shannon entropy in the position space decreases to the point where it becomes negative, indicating a greater location of the particle. However, in momentum space, Shannon entropy tends to increase, in such a way that the total entropy of the system increases. This indicates that the certainty in the location of the particle decreases because it is less localized. 

When we increase the energy level of the system, Shannon entropy increases, also increasing the uncertainty in the location of the particle. This indicates a delocalization of the particles as we increase the energy state of the system. Again, our analysis is done only for the three fundamental energy levels of the system, and in all of them, the BBM uncertainty relation is obeyed.

\begin{table}[ht!]
\centering
\begin{tabular}{|c|c|c|c|c|c|c|c|c|c|c|}\hline
\hline
$n$ & $\alpha$ & $r_0$ & $D_e$ & $B$ & $\Phi$ & $S_{r}$ & $S_{p}$ & $S_{r}+S_{p}$ & $1+\ln\pi$\\ \hline
\hline
  & 3/4 & 1 & 1 & 1 &  3/4  & 0.34703  & 2.11158  & 2.45862 & \\
0 & 3/4 & 2 & 1 & 1 &  3/4  & 0.49255  & 1.73543  & 2.22799 & 2.14473 \\
  & 3/4 & 3 & 1 & 1 &  3/4  & 0.53350  & 1.69360  & 2.22710 & \\ \hline
  & 3/4 & 1 & 1 & 1 &  3/4  & 0.34703  & 2.11158  & 2.45862 & \\
0 & 3/4 & 1 & 2 & 1 &  3/4  & 0.25081  & 2.32543  & 2.57625 & 2.14473 \\
  & 3/4 & 1 & 3 & 1 &  3/4  & 0.18197  & 2.42326  & 2.60523 & \\ 
\hline
  & 3/4 & 1 & 1 & 1 &  3/4  & 0.34703  & 2.11158  & 2.45862 & \\
0 & 3/4 & 1 & 1 & 2 &  3/4  & 0.12694  & 2.63157  & 2.75852 & 2.14473 \\
  & 3/4 & 1 & 1 & 3 &  3/4  & -0.04327 & 3.11973  & 3.07646 & \\   
  \hline
  & 3/4 & 1 & 1 & 1 &  3/4  & 0.34703  & 2.11158  & 2.45862 & \\
0 & 1/2 & 1 & 1 & 1 &  3/4  & 0.23843  & 2.30804  & 2.54648 & 2.14473 \\
  & 1/4 & 1 & 1 & 1 &  3/4  & -0.02167 & 2.96743  & 2.94576 & \\ \hline
  & 3/4 & 1 & 1 & 1 &  3/4  & 0.34703  & 2.11158  & 2.45862 & \\
0 & 3/4 & 1 & 1 & 1 &  1/2  & 0.34238  & 2.13971  & 2.48209 & 2.14473 \\
  & 3/4 & 1 & 1 & 1 &  1/4  & 0.33881  & 2.16386  & 2.50268 & \\ \hline
 \hline
  & 3/4 & 1 & 1 & 1 &  3/4  & 0.56364  & 2.68028  & 3.24393 & \\
1 & 3/4 & 2 & 1 & 1 &  3/4  & 0.72677  & 2.19688  & 2.92366 & 2.14473 \\
  & 3/4 & 3 & 1 & 1 &  3/4  & 0.77714  & 2.07871  & 2.85585 & \\ \hline
  & 3/4 & 1 & 1 & 1 &  3/4  & 0.56364  & 2.68028  & 3.24393 & \\
1 & 3/4 & 1 & 2 & 1 &  3/4  & 0.47613  & 2.85589  & 3.33204 & 2.14473 \\
  & 3/4 & 1 & 3 & 1 &  3/4  & 0.41254  & 3.02634  & 3.43889 & \\ 
\hline
  & 3/4 & 1 & 1 & 1 &  3/4  & 0.56364  & 2.68028  & 3.24393 & \\
1 & 3/4 & 1 & 1 & 2 &  3/4  & 0.34355  & 3.31826  & 3.66182 & 2.14473 \\
  & 3/4 & 1 & 1 & 3 &  3/4  & 0.17333  & 3.93621  & 4.10956 & \\   
  \hline
  & 3/4 & 1 & 1 & 1 &  3/4  & 0.56364  & 2.68028  & 3.24393 & \\
1 & 1/2 & 1 & 1 & 1 &  3/4  & 0.46107  & 2.91164  & 3.37272 & 2.14473 \\
  & 1/4 & 1 & 1 & 1 &  3/4  & 0.21528  & 3.67230  & 3.88758 & \\ \hline
  & 3/4 & 1 & 1 & 1 &  3/4  & 0.56364  & 2.68028  & 3.24393 & \\
1 & 3/4 & 1 & 1 & 1 &  1/2  & 0.55528  & 2.72988  & 3.28517 & 2.14473 \\
  & 3/4 & 1 & 1 & 1 &  1/4  & 0.54899  & 2.77108  & 3.32009 & \\ \hline
 \hline
  & 3/4 & 1 & 1 & 1 &  3/4  & 0.67760  & 3.08203  & 3.75964 & \\
2 & 3/4 & 2 & 1 & 1 &  3/4  & 0.85263  & 2.49365  & 3.34629 & 2.14473 \\
  & 3/4 & 3 & 1 & 1 &  3/4  & 0.90996  & 2.33945  & 3.24942 & \\ \hline
  & 3/4 & 1 & 1 & 1 &  3/4  & 0.67760  & 3.08203  & 3.75964 & \\
2 & 3/4 & 1 & 2 & 1 &  3/4  & 0.59582  & 3.22958  & 3.82540 & 2.14473 \\
  & 3/4 & 1 & 3 & 1 &  3/4  & 0.53582  & 3.44160  & 3.97743 & \\ 
\hline
  & 3/4 & 1 & 1 & 1 &  3/4  & 0.67760  & 3.08203  & 3.75964 & \\
2 & 3/4 & 1 & 1 & 2 &  3/4  & 0.45752  & 3.81845  & 4.27597 & 2.14473 \\
  & 3/4 & 1 & 1 & 3 &  3/4  & 0.28730  & 4.53177  & 4.81907 & \\   
  \hline
  & 3/4 & 1 & 1 & 1 &  3/4  & 0.67761  & 3.08203  & 3.75964 & \\
2 & 1/2 & 1 & 1 & 1 &  3/4  & 0.57897  & 3.30096  & 3.87993 & 2.14473 \\
  & 1/4 & 1 & 1 & 1 &  3/4  & 0.34313  & 4.15724  & 4.50037 & \\ \hline
  & 3/4 & 1 & 1 & 1 &  3/4  & 0.67761  & 3.08203  & 3.75964 & \\
2 & 3/4 & 1 & 1 & 1 &  1/2  & 0.66690  & 3.14788  & 3.81478 & 2.14473 \\
  & 3/4 & 1 & 1 & 1 &  1/4  & 0.65892  & 3.17998  & 3.83890 & \\ \hline
\end{tabular}\\
\caption{Numerical result of Shannon entropy for pseudoharmonic and shifted pseudoharmonic potential ($l=0$).\label{tab2}}
\end{table}

\vspace{0.1cm}
{\bf Case D}
\vspace{0.1cm}

Finally, for the last case where we consider an inverse square potential, when we increase the value of the potential parameter ($b$), the less localized the particle will be in the position space. This is evident in Tab.\ref{tab3} where information entropy increases as $b$ increases. In position space, the particle becomes more localized. The sum of entropies tends to decrease as $b$ increases, i.e., the more localized the particle becomes (the smaller the uncertainty in the particle's location). The opposite happens when we increase the influence of the magnetic field $B$, where the entropy increases, indicating a smaller location, i.e., greater will be the uncertainties in the location of the particle.

When we consider $\alpha,\Phi\rightarrow1$ the entropic information measures will be larger in the position space and smaller they will be in the momentum space. The total entropy of the system decreases, which indicates a greater location of the particle, i.e., the uncertainty in the location of the particle in the system will be smaller. Furthermore, when we increase the system's energy levels, the greater the uncertainty of the particle's location, i.e., the less localized the particle will be. We only consider the three lowest energy levels, and in all of them, the BBM relationship is satisfied.

\begin{table}[ht!]
\centering
\begin{tabular}{|c|c|c|c|c|c|c|c|c|}\hline
\hline
$n$ & $\alpha$ & $b$ & $B$ & $\Phi$ & $S_{r}$ & $S_{p}$ & $S_{r}+S_{p}$ & $1+\ln\pi$\\ \hline
\hline
  & 3/4 & 1 & 1 &  3/4  & 0.53548  & 1.74891  & 2.28441 & \\
0 & 3/4 & 2 & 1 &  3/4  & 0.54548  & 1.72255  & 2.26804 & 2.14473 \\
  & 3/4 & 3 & 1 &  3/4  & 0.55094  & 1.67033  & 2.22128 & \\ \hline
  & 3/4 & 1 & 1 &  3/4  & 0.53548  & 1.74891  & 2.28441 & \\
0 & 3/4 & 1 & 2 &  3/4  & 0.18890  & 2.47334  & 2.66225 & 2.14473 \\
  & 3/4 & 1 & 3 &  3/4  & -0.01382 & 3.02921  & 3.01539 & \\ \hline
  & 3/4 & 1 & 1 &  3/4  & 0.53548  & 1.74891  & 2.28441 & \\
0 & 1/2 & 1 & 1 &  3/4  & 0.33980  & 2.08185  & 2.42166 & 2.14473 \\
  & 1/4 & 1 & 1 &  3/4  & 0.00777  & 2.88067  & 2.88845 & \\ \hline
  & 3/4 & 1 & 1 &  3/4  & 0.53548  & 1.74891  & 2.28441 & \\
0 & 3/4 & 1 & 1 &  1/2  & 0.53082  & 1.77221  & 2.30304 & 2.14473 \\
  & 3/4 & 1 & 1 &  1/4  & 0.52725  & 1.79221  & 2.31948 & \\ \hline
  \hline
  & 3/4 & 1 & 1 &  3/4  & 0.75208  & 2.19577  & 2.94787 & \\
1 & 3/4 & 2 & 1 &  3/4  & 0.77080  & 2.11909  & 2.88989 & 2.14473 \\
  & 3/4 & 3 & 1 &  3/4  & 0.78151  & 2.08346  & 2.86498 & \\ \hline
  & 3/4 & 1 & 1 &  3/4  & 0.75208  & 2.19577  & 2.94787 & \\
1 & 3/4 & 1 & 2 &  3/4  & 0.40551  & 3.11764  & 3.52316 & 2.14473 \\
  & 3/4 & 1 & 3 &  3/4  & 0.20278  & 3.82305  & 4.02584 & \\ \hline
  & 3/4 & 1 & 1 &  3/4  & 0.75208  & 2.19577  & 2.94787 & \\
1 & 1/2 & 1 & 1 &  3/4  & 0.56244  & 2.62861  & 3.19105 & 2.14473 \\
  & 1/4 & 1 & 1 &  3/4  & 0.24473  & 3.56620  & 3.81094 & \\ \hline
  & 3/4 & 1 & 1 &  3/4  & 0.75208  & 2.19577  & 2.94787 & \\
1 & 3/4 & 1 & 1 &  1/2  & 0.74373  & 2.26101  & 3.00475 & 2.14473 \\
  & 3/4 & 1 & 1 &  1/4  & 0.73744  & 2.29512  & 3.03257 & \\ \hline
    \hline
  & 3/4 & 1 & 1 &  3/4  & 0.86604  & 2.55266  & 3.41871 & \\
2 & 3/4 & 2 & 1 &  3/4  & 0.89048  & 2.42533  & 3.31582 & 2.14473 \\
  & 3/4 & 3 & 1 &  3/4  & 0.90480  & 2.37398  & 3.27879 & \\ \hline
  & 3/4 & 1 & 1 &  3/4  & 0.86604  & 2.55266  & 3.41871 & \\
2 & 3/4 & 1 & 2 &  3/4  & 0.51947  & 3.58578  & 4.27597 & 2.14473 \\
  & 3/4 & 1 & 3 &  3/4  & 0.31674  & 4.39959  & 4.71634 & \\ \hline
  & 3/4 & 1 & 1 &  3/4  & 0.86604  & 2.55266  & 3.41871 & \\
2 & 1/2 & 1 & 1 &  3/4  & 0.68033  & 3.01140  & 3.69173 & 2.14473 \\
  & 1/4 & 1 & 1 &  3/4  & 0.37257  & 4.03574  & 4.40832 & \\ \hline
  & 3/4 & 1 & 1 &  3/4  & 0.86604  & 2.55266  & 3.41871 & \\
2 & 3/4 & 1 & 1 &  1/2  & 0.85533  & 2.60722  & 3.46256 & 2.14473 \\
  & 3/4 & 1 & 1 &  1/4  & 0.84735  & 2.65187  & 3.49923 & \\ \hline
\end{tabular}\\
\caption{Numerical result of Shannon entropy for inverse square potential ($l=0$).\label{tab3}}
\end{table}

\section{Conclusions}
\label{sec5}

In this research, we have investigated the behavior of charged particles confined by a flux field with a potential in the background of a cosmic dislocation. Numerous approaches have been employed by various authors to obtain exact and approximate eigenvalue solutions for wave equations with different types of potentials. In this study, we have opted for the parametric NU-method, a technique that has been successfully utilized by many researchers in previous works. We derived the radial equation specific to our system and analytically solved it using the NU-method.

As a result of our analysis, we have determined the energy levels using Eq. (\ref{13}) and obtained the normalized radial wave function using Eq. (\ref{14}) for non-relativistic particles confined by both the AB-flux and magnetic fields. It is evident that these energy levels and the wave function are influenced by the presence of a topological defect represented by the parameter $\alpha$, as well as the magnetic and quantum flux fields in conjunction with the anharmonic oscillator potential. Consequently, the energy eigenvalues deviate from the results in flat space and break the degeneracy. In the limit as $\alpha \to 1$, Eqs. (\ref{13}) and (\ref{14}) converge to the Landau solution with this specific potential. To provide visual insights into our findings, we have created graphical representations of the energy levels (Fig.\ref{fig:2}) and the normalized radial wave function (Fig.\ref{fig:3}) while varying the values of the different parameters. In sub-section \ref{sec2.1}, we employed the derived eigenvalue solution to explore various molecular potential models, including the harmonic oscillator potential (Case A), pseudoharmonic potential (Case B), shifted pseudoharmonic potential (Case C), and inverse square potential (Case D). We presented the eigenvalue solutions for each of these cases. Our analysis clearly revealed that these eigenvalue solutions are subject to the influence of the topological defect, as well as the magnetic and quantum flux fields.

In Section \ref{sec3}, our focus shifted towards the analysis of thermodynamic properties of the quantum system at a finite temperature, specifically for $T \neq 0$. We began by calculating the partition function $Z(\beta)$ using the energy expression given by Eq. (\ref{13}). It became evident that this partition function is dependent on several factors, including the topological defects associated with the cosmic string (represented by the parameter $\alpha$), the potential strengths $a$ and $b$, as well as the presence of magnetic and quantum flux fields $(B, \Phi_{AB})$. With the partition function in hand, we proceeded to determine various thermodynamic functions, including the vibrational free energy, mean free energy, specific heat capacity, and entropy of the quantum system. Our analysis revealed that these thermodynamic quantities are significantly influenced by the topological defects characterized by the parameter $\alpha$, the potential parameters $a$ and $b$, as well as the magnetic and quantum flux fields. To provide a visual representation of these findings, we generated several graphs illustrating the behavior of these thermodynamic quantities with respect to the topological defect parameter $\alpha$ for different values of the magnetic field and the quantum flux, as shown in Figs.\ref{fig:4} to \ref{fig:7}.  Furthermore, we conducted an in-depth examination of the magnetic properties of the quantum system under two distinct temperature conditions: zero temperature ($T=0$) and finite temperature ($T \neq 0$). We thoroughly analyzed the effects of various factors on these magnetic properties. Our analysis involved the computation of persistent currents, denoted as $I_{n,l}$, along with the magnetization and magnetic susceptibility. Our results demonstrated that these magnetic properties are significantly influenced by several factors, including the topological defect associated with the cosmic string (represented by the parameter $\alpha$), the potential parameters $a$ and $b$ of the anharmonic oscillator potential, as well as the presence of magnetic and quantum flux fields ($B, \Phi_{AB}$).

In section \ref{sec4}, we studied the entropic information of the quantum system under investigation where, cosmic disclinations, potential and the external magnetic and quantum flux fields effects are considered. The quantum information via Shannon entropy showed us that when we increase the influence of the less localized magnetic field the particle will be, this is because the uncertainties in the location of the particle will be greater. The same happens when we reduce the influence of the topological defect and the quantum flow. Furthermore, the parameters that control the shape of the potential also affect the location of the particle. In case A when we increase the value of the angular oscillation frequency ($\omega$), we decrease the certainty in the location of the particle (delocalization). In cases B and C, when we decrease the value of the equilibrium inter-nuclear separation ($r_0$) and when we increase the value of the dissociation energy ($D_e$), the greater the uncertainties in the location of the particle will be. In case D, it was possible to observe that when we increase the value of $b$, the greater the certainty in the location of the particle. Furthermore, as we increase the energy level of the system, the particle is less localized, i.e., the greater the uncertainties in location will be. 

\section*{Conflict of Interest}

Authors declare(s) no such competing interests.

\section*{Data Availability Statement}

All generated data are included in this manuscript.

\section*{Funding Statement}

No fund has received for this study.

\end{document}